\documentclass[sigconf]{acmart}
\usepackage{multirow}
\usepackage{longtable}
\usepackage{multirow}
\usepackage{longtable}
\usepackage{setspace}
\usepackage{natbib}
\usepackage{float}
\usepackage{dblfloatfix}
\usepackage{array}
\usepackage{enumitem}
\usepackage[normalem]{ulem}
\usepackage[final]{pdfpages}
\usepackage{ragged2e}
\newcommand{\ignore}[1]{}

\newcolumntype{P}[1]{>{\RaggedRight\arraybackslash}p{#1}}
      \makeatletter
\def\@fnsymbol#1{\ensuremath{\ifcase#1\or \dagger\or \ddagger\or
   \mathsection\or \mathparagraph\or \|\or **\or \dagger\dagger
   \or \ddagger\ddagger \else\@ctrerr\fi}}
    \makeatother
    
\AtBeginDocument{%
  \providecommand\BibTeX{{%
    \normalfont B\kern-0.5em{\scshape i\kern-0.25em b}\kern-0.8em\TeX}}}

\copyrightyear{2023}
\acmYear{2023}
\setcopyright{acmlicensed}\acmConference[CSCW '23 Companion]{Computer Supported Cooperative Work and Social Computing}{October 14--18, 2023}{Minneapolis, MN, USA}
\acmBooktitle{Computer Supported Cooperative Work and Social Computing (CSCW '23 Companion), October 14--18, 2023, Minneapolis, MN, USA}
\acmPrice{15.00}
\acmDOI{10.1145/3584931.3607014}
\acmISBN{979-8-4007-0129-0/23/10}

\begin{document}
\title[Perspectives from Naive Participants and Social Scientists on\\Addressing Embodiment in a Virtual Cyberball Task]{Perspectives from Naive Participants and Social Scientists on Addressing Embodiment in a Virtual Cyberball Task}


\author{Tao Long}
\authornote{Currently affiliated with Columbia University. This work was his undergrad research honors thesis at Cornell: \textit{Designing a Virtual Cyberball Task for Social Science Research}.\vspace{1pt}}
\email{tl468@cornell.edu}
\affiliation{%
  \institution{Cornell University}
  \streetaddress{}
  \city{Ithaca}
  \state{New York}
  \country{USA}
  \postcode{10027}
}

\author{Swati Pandita}
\authornote{Currently affiliated with California Institute of Technology.}
\email{sp2333@cornell.edu}
\affiliation{%
  \institution{Cornell University}
  \streetaddress{}
\city{Ithaca}
  \state{New York}
  \country{USA}
  \postcode{10027}
}
\author{Andrea Stevenson Won}
\email{asw248@cornell.edu}
\affiliation{%
  \institution{Cornell University}
  \streetaddress{}
  \city{Ithaca}
  \state{New York}
  \country{USA}
  \postcode{10027}
}

\renewcommand{\shortauthors}{Long et al.}

\begin{abstract}
We describe the design of an immersive virtual Cyberball task that included avatar customization, and user feedback on this design. We first created a prototype of an avatar customization template and added it to a Cyberball prototype built in the Unity3D game engine. Then, we conducted in-depth user testing and feedback sessions with 15 Cyberball stakeholders: five naive participants with no prior knowledge of Cyberball and ten experienced researchers with extensive experience using the Cyberball paradigm. We report the divergent perspectives of the two groups on the following design insights; designing for intuitive use, inclusivity, and realistic experiences versus minimalism. Participant responses shed light on how system design problems may contribute to or perpetuate negative experiences when customizing avatars. They also demonstrate the value of considering multiple stakeholders' feedback in the design process for virtual reality, presenting a more comprehensive view in designing future Cyberball prototypes and interactive systems for social science research.

\end{abstract}

\ccsdesc[500]{Human-centered computing~Empirical studies in HCI}
\ccsdesc[300]{Human-centered computing~Virtual reality}

\keywords{virtual reality; cyberball task; social science paradigm; walkthrough; avatar customization; qualitative research; semi-structured interviews; user experience; multi-stakeholders; inclusive design; games}

\begin{teaserfigure}
\vspace{-3pt}
\minipage{0.33\textwidth}
  \includegraphics[width=\linewidth]{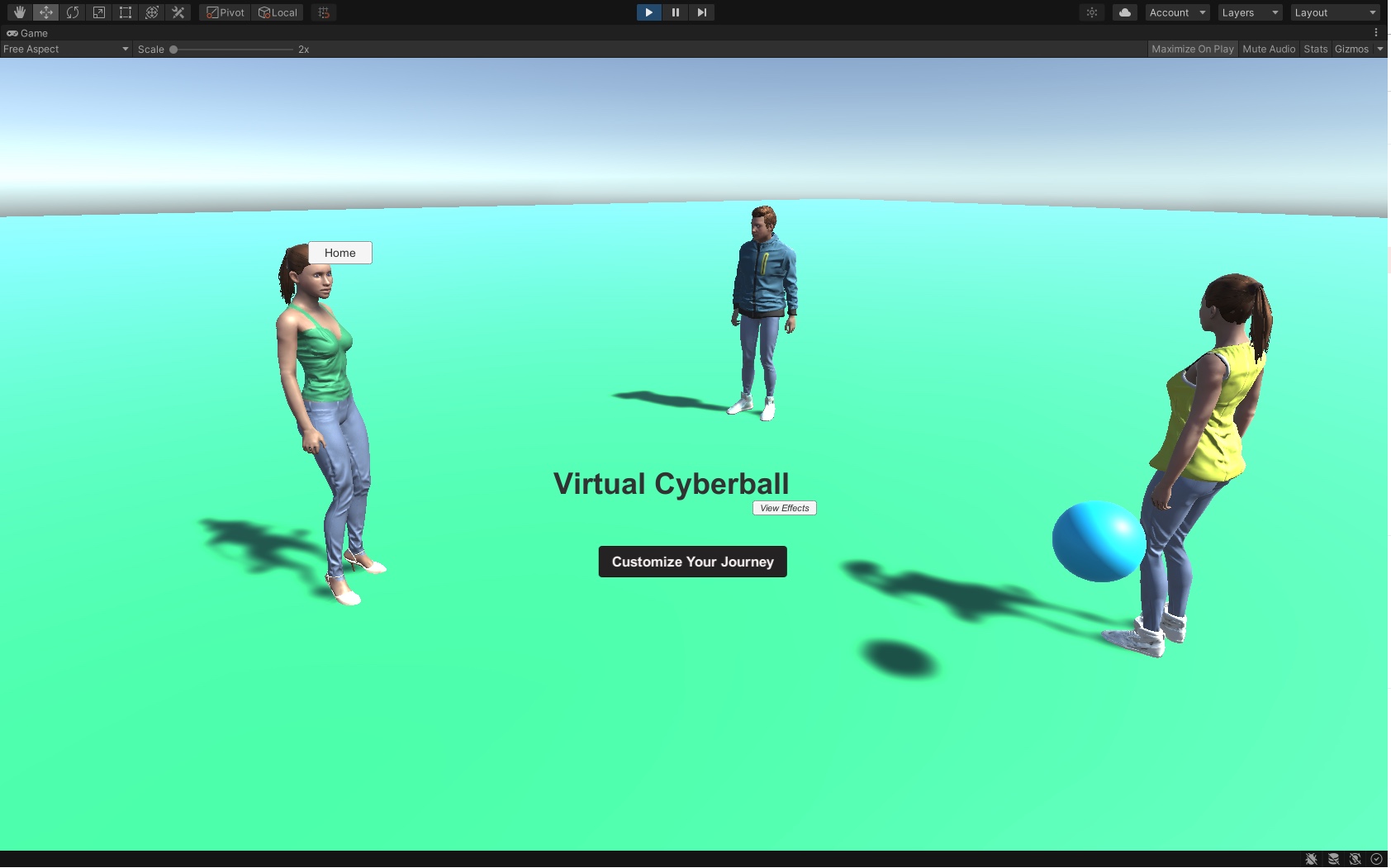}
   \Description[A screenshot for Scene 1]{Scene 1 contains a line of Welcome Message, an illustration of three avatars and a ball on a large green plane, and a Start button for users to start the game.}
  \caption{Scene 1: Welcome Message}
\endminipage\hfill
\minipage{0.33\textwidth}
  \includegraphics[width=0.995\linewidth]{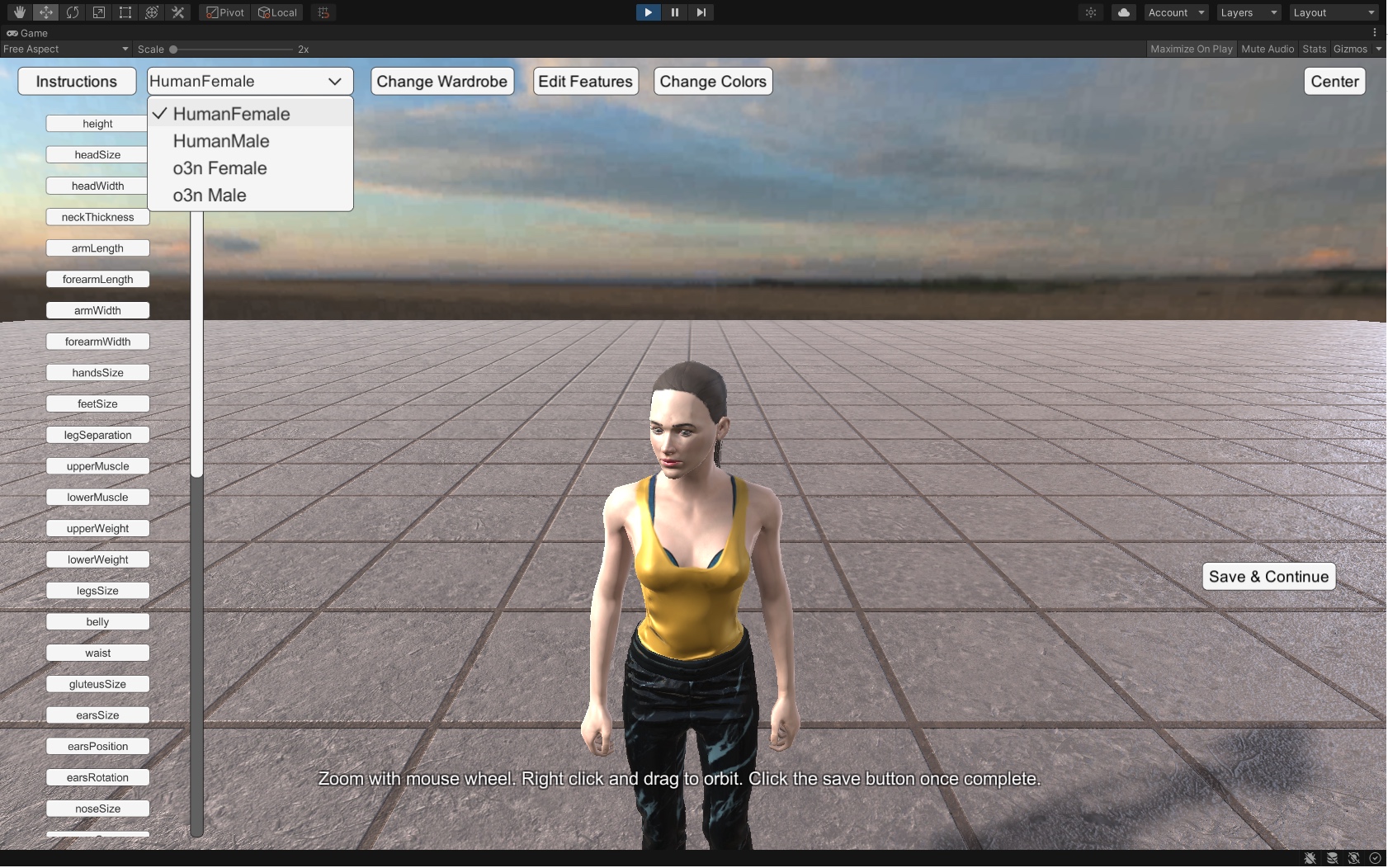}
  \Description[A screenshot for Scene 2]{Scene 2 offers a list of customization options for users to edit their avatar: body and facial features, hair and skin color, and clothing. Users could slide the bar or click different buttons at the top of the screen to customize their avatar. They could also see the effects on their avatar, who stands on a brown plane at the bottom of the screen. There is a "Save and Next" button at the right of the screen for them to save their customization after they are done.}
  \caption{Scene 2: Avatar Customization}
\endminipage\hfill
\minipage{0.33\textwidth}%
  \includegraphics[width=0.995\linewidth]{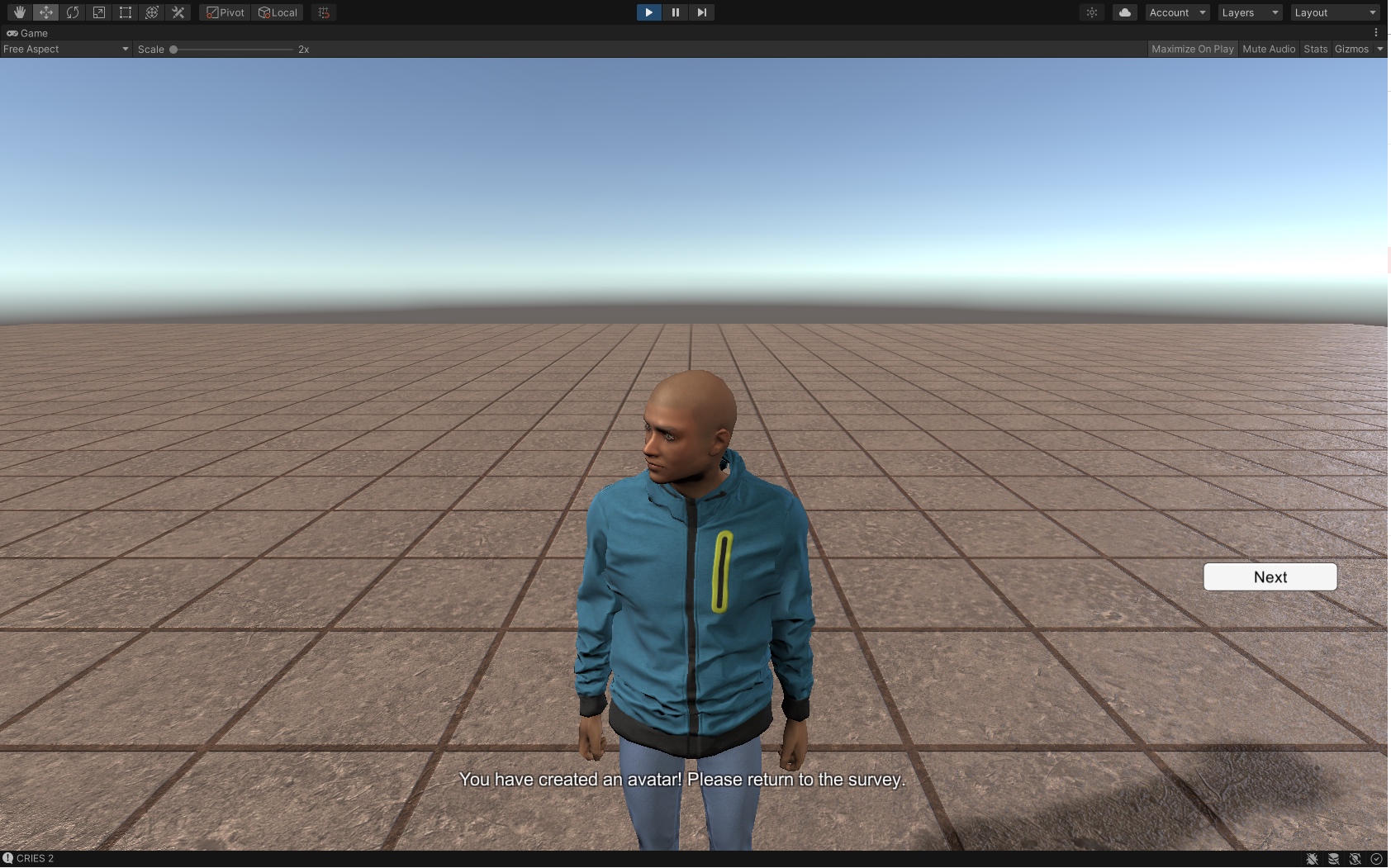}
    \Description[A screenshot for Scene 3]{Scene 3 provides users a closeup view of their final avatar. Users' avatar stands on a large brown plane and users could click the "Next" button at the right side of the screen to start the game.}
  \caption{Scene 3: Close-up of Final Avatar}
\endminipage\hfill
\newline
\newline
\minipage{0.33\textwidth}
  \includegraphics[width=0.995\linewidth]{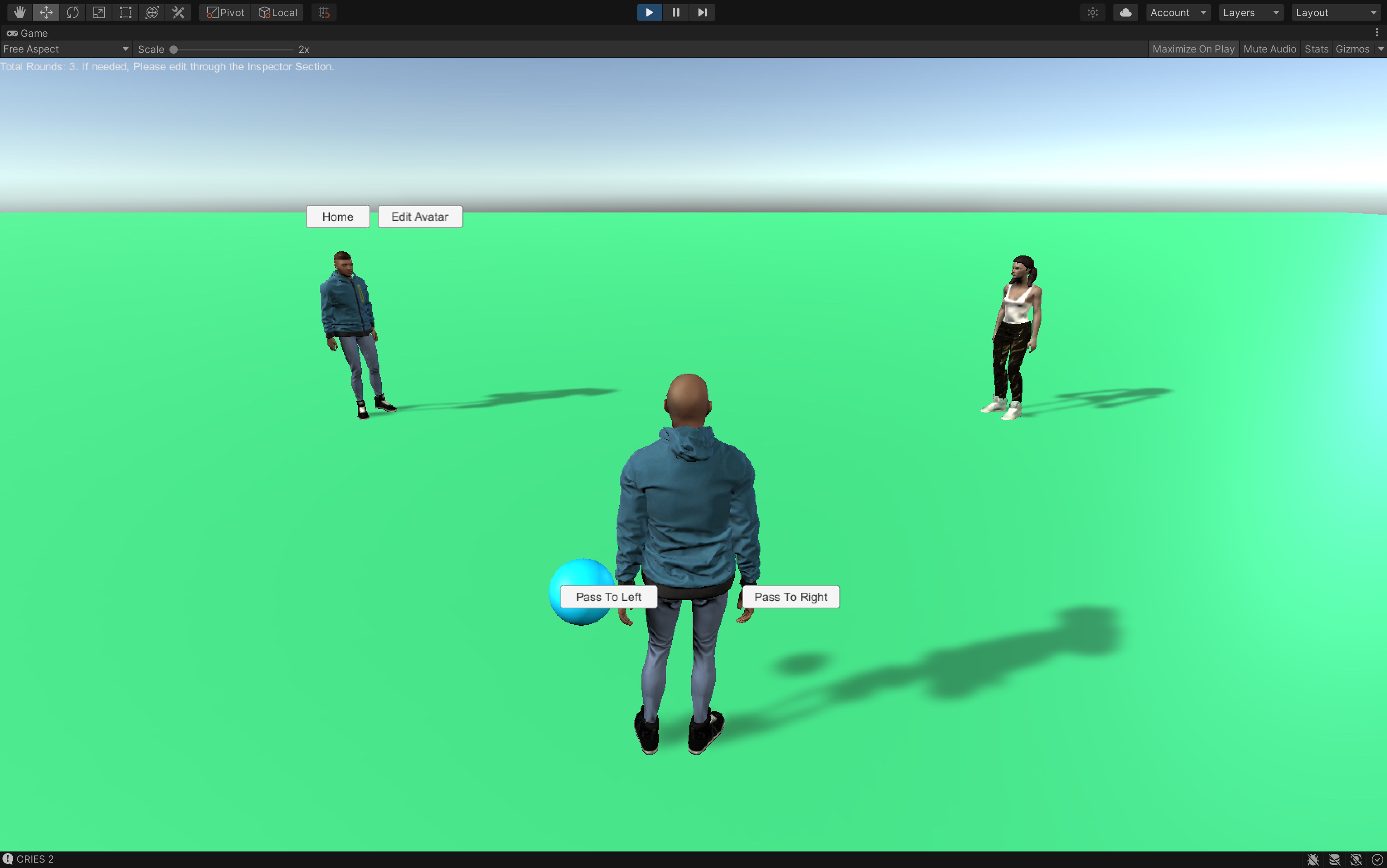}
    \Description[A screenshot for Scene 4]{Scene 4 is the exclusive game scene. Users' customized avatar and two computer agents are facing each other on a large green plane. There is a blue ball next to the left hand of the users' avatar, and there are two buttons next to them: "Pass to left" button, and "Pass to right" button. users could click on of it and the ball will be passed to agent who is either on the left side or on the right side. This screenshot only shows the third-person point of view, so users could see the full body of their customized avatar.}
  \caption{Scene 4: Exclusive Game \textmd{\\(This shows the 3rd person POV version)}}
\endminipage\hfill
\minipage{0.33\textwidth}
  \includegraphics[width=0.995\linewidth]{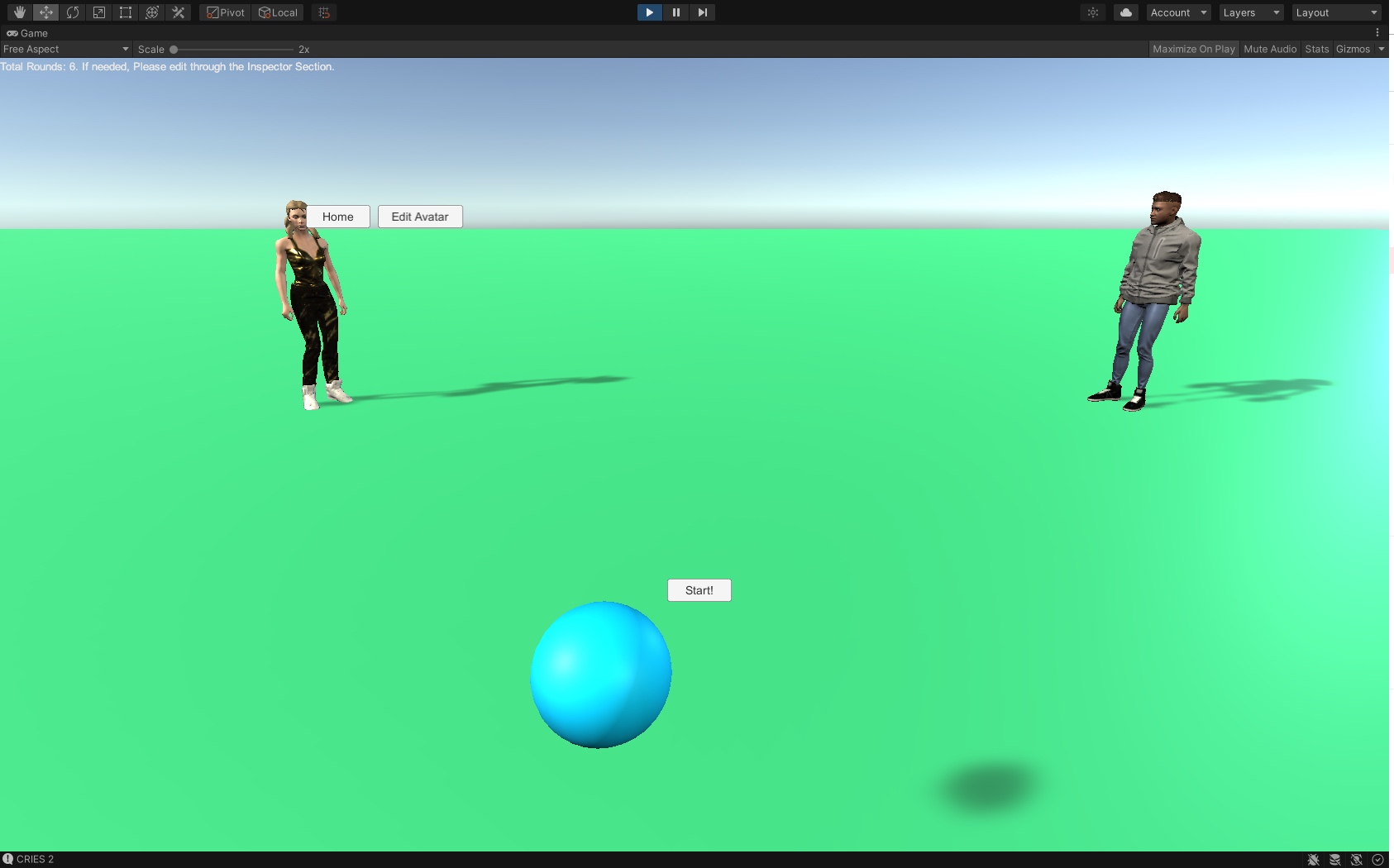}
   \Description[A screenshot for Scene 5]{Scene 5 is the inclusive game scene. As this screenshot only shows the first-person point of view, users could only see two computer agents are facing them on a large green plane.There is a blue ball right in front of the user, and there are two buttons next to them: "Pass to left" button, and "Pass to right" button. users could click on of it and the ball will be passed to agent who is either on the left side or on the right side.}
   \caption{Scene 5: Inclusive Game \textmd{\\(This shows the 1st person POV version)}}
\endminipage\hfill
\minipage{0.33\textwidth}%
  \includegraphics[width=0.995\linewidth]{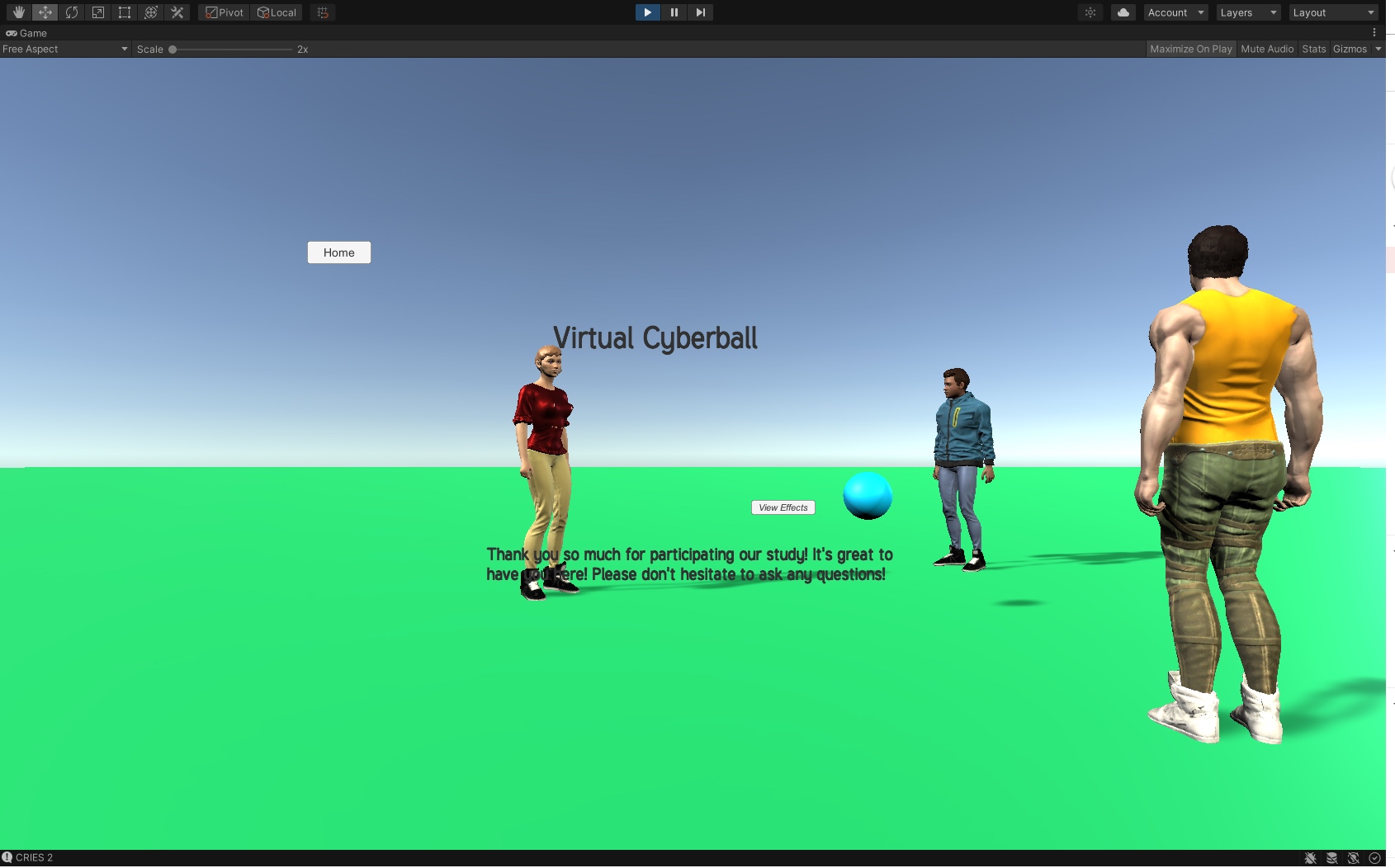}
   \Description[A screenshot for Scene 6]{Scene 6 contains a line of Thank You Message, an illustration of three avatars and a ball on a large green plane, and a Home button to return to the front page.}
  \caption{Scene 6: Finish Message\newline}
\endminipage
\vspace{3pt}
\Description[Screenshots of the virtual Cyberball system by scenes]{Screenshots of the virtual Cyberball system by scenes}
\end{teaserfigure}

\maketitle
\section{Introduction}

\begin{figure}[h]
 \begin{center}
 \vspace{-11pt}
\includegraphics[width=0.38\textwidth]{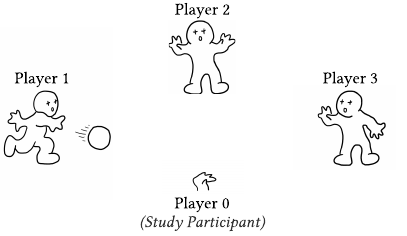}
\Description[a game between four players, who are throwing balls at each other]{This is an illustration of Cyberball, an online ball-tossing game. They each have a small character on the screen and pass one ball to each other throughout the session.}
\caption{2D Cyberball design, following \cite{pryor_influence_2013}}
\label{fig:2d}
 \end{center}
  \vspace{-15pt}
\end{figure}

Cyberball is a popular social science paradigm for testing the effects of social inclusion and exclusion \cite{williams_cyberball_2006, 10.3389/fnhum.2014.00935}. As Figure \ref{fig:2d} shows, participants can be either included (tossed the ball an equal amount of times) or excluded (ignored by others) during the game. 

Currently, the fifth version of the Cyberball paradigm \cite{2019} is widely used in research on mood, self-esteem, social pain, internet ostracism, and cyberbullying \cite{boyes_having_2009, zadro_how_2004, wudarczyk_chemosensory_2015, kawamoto_cognitive_2013, le_perceived_2020}. In efforts to improve task saliency, researchers have adapted and modified the original paradigm by creating online versions \cite{williams_cyberostracism_2000, zadro_how_2004}, eliminating or adding rounds of the games \cite{dorn_behavioral_2014}, displaying on different devices \cite{kano_factors_2019}, and integrating measures such as fMRI \cite{sebastian_developmental_2011}.

In 2012, a virtual reality (VR) version of Cyberball \cite{kassner_virtually_2012} was developed to give better systematic control, similar to other methods, while also allowing for richer in-game social experiences \cite{kassner_virtually_2012, wirth_methods_2016, kothgassner_does_2020}. While the authors successfully replicated the Cyberball paradigm into a more immersive experience, the environment intentionally lacked a critical feature: avatar customization (AC). The rationale for excluding AC was to isolate the effects of social ostracism through play like the original paradigm \cite{kassner_virtually_2012}. Therefore, limiting other drivers of ostracism such as social identity~\cite{goodwin2010psychological}, which can be manipulated through player avatar customization \cite{pandita_humphreys_won_2023}, was imperative. However, as previous AC and embodiment research~\cite{kang2020my, embodimentnew} have shown, AC can increase player identification, thus bolstering the effects of the paradigm. 
Furthermore, it can increase user engagement and Cyberball's ecological validity \cite{6632611, 10.1145/3474675}. 

To update this tool by making it more \textit{\textbf{engaging}} for game experience and more \textit{\textbf{useful}} for scientific research, we developed an immersive Cyberball prototype\footnote{GitHub repository link: \url{https://github.com/iamtaolong/virtual-cyberball}} 
that included the ability to customize avatars. We anticipated that two groups of stakeholders would be critical in understanding how the VR Cyberball experience could be improved. One key group was \textit{researchers} who actively use the paradigm to gather data to answer their research questions. The second group consisted of \textit{naive participants} to represent potential study participants: they have no prior knowledge of or experience with the Cyberball paradigm. 
We conducted a user study with 15 participants
and collected user feedback from both groups. 

Our research questions were as follows:
\begin{itemize} 
    \item[\textit{\textbf{RQ1:}}] What design considerations or needs are \textit{shared} or \textit{differ} between experienced researchers and naive participants?
    \item[\textit{\textbf{RQ2:}}] What recommendations can we make for future VR Cyberball and social science paradigm design?
\end{itemize}

\section{Prototype Design and Development}
The virtual environment was built with the Unity 3D game engine. It was first deployed as a WebGL environment that studied the effects of avatar customization on mood \cite{panditaDissertation} with the Cyberball paradigm described in \cite{kassner_virtually_2012}. It was then modified for immersive VR for the current study. A key difference between the version Kassner and colleagues developed and ours was the addition of an avatar customization feature, using a representative off-the-shelf customization menu \cite{uma2021}. We added this in order to examine how different participant groups experienced avatar representation. 

Scene 1 welcomed participants. In Scene 2, users were presented with the default template avatar (a Caucasian female) and customized their avatar through buttons and sliders. 
Editable features included body and facial features, body and head size, hair style and color, and clothing. After clicking the \textit{Save \& Continue} button, each user's customization history was automatically downloaded as a .csv file to the desktop. For Scene 3, users were presented with a close-up view of their finished avatar. For Scene 4 and Scene 5, the users entered the exclusion and inclusion games. Finally, Scene 6 ended with a message of thanks for participation. To deploy the virtual environment to a Quest 2 headset, we replaced the keyboard input with a device-based and action-based locomotion system.

\section{User Testing}
\subsection{Methods and Participants} 
A total of ten researchers (R1-R10) and five naive participants (P1-P5) were recruited for the study from March to July 2022. We conducted user evaluation sessions using either a VR headset in a physical lab or a screen-shared walkthrough demo via Zoom. Lastly, participants were interviewed about their experience and insights on the environment and game design. This research was determined to be exempt after consultation with the university IRB board.
We used purposeful sampling to recruit ten experienced researchers across the Americas and Europe who had conducted Cyberball-based research from the perspectives of psychology, human development, nutritional science, and sociology. After finding publications related to Cyberball on Google Scholar, we recruited the authors through inquiry emails. 
We recruited from personal sampling five naive participants from a U. S. university (two men, three women), undergraduate students with no prior experience with Cyberball, ages 20-23. Four identified as Asian and one as white. Three had prior experience with VR, and all were familiar with customizing avatars from mobile or video game experiences. 

\subsection{Study Flow}
Researchers (R1-R10) were first asked about their challenges and concerns in adapting Cyberball to their research field. As all ten researchers were remote, we utilized the walkthrough method \cite{doi:10.1177/1461444816675438} via Zoom. The first author demonstrated the environment in Unity for them. During the demo, the researchers could also share where they wanted to go and which buttons they wanted to click. Then, researchers were asked to share their thoughts on the system design and whether it fits their needs. Each testing and interview session with researchers took around thirty minutes, and participants verbally consented prior to the interview and audio recording.

Naive participants (P1-P5) tried out the system and played with the environment inside the Oculus Quest 2 head-mounted display and Unity during in-person evaluation sessions. Then, they reported their experiences in each scene, emphasizing what was enjoyable and what needed improvements. We asked open-ended questions to understand their detailed experiences with the design of avatars, agents, background environment, and ball-tossing game. Lastly, participants were asked about potential system improvements and demographics. Each interview took approximately 20 minutes, and the first author conducted interviews and note-taking.

We analyzed the data from the interview sessions with both researchers and naive participants using thematic analysis with an inductive approach \cite{pan2018and, braun_clarke_2012}. After reading through the notes and the recording transcripts, we established three significant themes based on their experiences with Cyberball. They are 1) design for intuitive use, 2) design for inclusivity, and 3) design for realistic experiences. Then, we analyzed our interview data for the ten researchers and coded all the related responses into the themes, thus establishing a comparison between the two groups of stakeholders: the third theme became 3) design for realistic experiences or minimalism.

\section{Evaluation and Feedback}
In fifteen user interviews, all stakeholders expressed favorable comments about the virtual environment design. 
Among them, all five naive participants noted "the importance of moving [research paradigm] into an immersed environment" (P1) and commented on the detailed customization available. One naive participant mentioned that "the options could make people feel included and excited" (P4). The ten researchers agreed on the relevance and applicability of the virtual Cyberball paradigm to social science research. Eight found it especially applicable to their specific research topics. Both R1, R3, and R6 mentioned that this customization scene looks "awesome" and "powerful," and "[showing] real human features that you could identify groups [and] group belonging" (R6). 
R6 expressed that the possibilities of research questions could be largely expanded by adding complexities to a simple system. Also, some researchers recommended bringing the design to a broader audience and open-sourcing it: "I would highly recommend that you submit this to 
[Conference Name Redacted]..." (R3).
Below, we describe how participants agreed or differed on each theme. 

\subsection{Design for Intuitive Use}

\subsubsection*{Enhancing Embodiment Experiences Through Design}
Both naive participants and researchers mentioned the importance of adopting more visual cues to improve users' experiences of avatar embodiment. 
For example, regarding the avatar's height: "It's a little hard to tell how tall the avatar is since there isn't any reference object next to it. I am just moving the slider bar, but I cannot see any noticeable change." (P1) Also, P2 commented that it was hard to see the avatar's whole body or specific body parts, limiting their understanding of the avatar. Interestingly, P2 modified the height of their avatar by moving the slider towards the far right. Thus, after entering the inclusive game scene and comparing it with the other two agents inside the game, they were shocked: "Wow, why is [my avatar] this tall? Or these two [agents] are just too short?" (P2)
Similarly, researchers R1 and R4, who have experience developing Cyberball, mentioned that it would be interesting for users to name and give a short description of their avatar: "would be nice [to give] the person's name and their age [to the avatar... so] that way you can create in-groups and out-groups things" (R1). 

\subsubsection*{Leveraging Previous Experience with Technology}
Two naive participants mentioned their experiences with the avatar customization interface in other game designs. 
Many avatar customization scenes in mobile games allow users to drag the images of the clothing or hairstyles onto the avatar to see what the options look like. P1 said: "It's hard to click on each option and then look at the changes one by one" (P1). Also, P5 mentioned: "[with all the clothing options laying over the avatar] the users could just scroll the options from left to right over the avatars" (P5). Leveraging users' experiences with other games can aid their interactions with the interface. However, many researchers discussed how technology unfamiliarity might affect the experiences of senior users. For example, R2 mentioned that the instruction and user flow need be more intuitive and straightforward so all age groups know how to play it easily.

\subsection{Design for Inclusivity}
\subsubsection*{Race}
We intentionally used the most common commercially available customization menu to understand how different user groups experienced bias present in the avatar preset packages. As previous works indicate, though users can modify skin colors and detailed facial features, customization was challenging \cite{pace2009socially,kafai2010your, 10.1145/3270316.3270599}. As R1 stated, after shifting the skin color, the avatar's facial details and appearances still looked Caucasian.
Thus, a successful virtual paradigm needs "to make sure... [for] people's racial and ethnic identity and appearance... everybody feels like they can represent themselves" (R4). Plus, for a social science paradigm that examines social exclusion, researchers may need to represent a range of different age groups or racial/ethnic groups (R3). 

\subsubsection*{Age}
Five researchers mentioned this sub-theme as they had previous experiences applying the Cyberball paradigm to different age groups, similar to \cite{genderage, olderad, eeverone}. They foresaw problems for users embodying avatars that are significantly older or younger than the participants themselves. R2 stated, "you should be able to choose young, middle-aged, and old: 20s, 30s, 40s, 50s, 60s," and R5 also mentioned the importance of including various body types. Since their previous research with Cyberball focused on how different age groups handle social rejections, R1 and R2 mentioned the potential risks of using a young avatar for seniors: "we know that there are age differences in decision making. [For this scenario,] they're like deciding for a younger person, not themselves, so that they may change their mindsets" (R2). 

\subsubsection*{Gender}
Similar to racial biases, researchers like R7 and R8 mentioned stereotypes associated with the female avatar presets. As earlier literature \cite{genderage, genderbelt} shared, R7 shared that "[the] first thing I noticed was that the female-body avatars were wearing much less clothing" (R7), bringing "a potential risk of stereotype or threat, especially if you're trying to use this with underage participants like vulnerable populations" (R7). Also, R8 pointed out that many other gender-related threats are still embedded inside the VR preset packages. For example, purchasing clothing sets for female avatars costs more than for male avatars while modifying the presets. 

\subsubsection*{Neurodiversity or Diverse Physical Abilities}
Similar to \cite{10.1145/3517428.3544829}, one researcher mentioned that participants with physical difficulties might struggle to access the virtual paradigm, especially with hand controllers. Thus, besides an immersive version inside VR headsets, the researcher suggested including a more "user-friendly" way to interact with the system (R3). Also, regarding the risks of motion sickness, R2 and P1 mentioned, "[the agents] are swaying back and forth, thus making me dizzy... VR headsets always make me a bit nauseous, and just seeing the screen makes me crazy" (R2). Thus, for future application in social science research, this immersed Cyberball paradigm should still offer multiple versions via WebGL, customer headsets, and other accessibility-oriented designs to meet the needs of special user groups.

\subsection{Design for Realistic Experiences or Minimalism}
\subsubsection*{Avatar Customization Design}
The two groups each held different opinions on the desirable level of detail for the virtual environment. Most naive participants wanted the environment to be more "fun," especially regarding avatar customization. For example, P1 wanted a more exaggerated and random avatar design for the agents since it can be more interesting and exciting to play with "monsters." However, researchers wanted avatar customization to be straightforward and "minimal" (R1) for users to understand quickly. R4 mentioned that there should be fewer choices for avatar customization since "users may easily skip over" many options that are stored too deep in the menus. R3 and R4 shared that they did not want to give the participant too much power to customize the game by customizing anything other than their embodied avatars. They mention that "there is too many options for now... [which is] overwhelming and people might skip over it" (R4).

\subsubsection*{Agent-Avatar Design}
Some naive participants were not interested in the agent-avatars (the other "players" that were controlled by the computer). P4 and P5 stated they did not want to spend time or attention on agents' customization. Instead, following their prior experience with avatar customization, they wanted more focus on their avatar. However, some researchers discussed the need to customize agents.
For R9 and R10, two researchers who are conducting Cyberball-related research in Eastern Europe mentioned it is essential to make the agent design realistic and background-specific. They shared that applying the same race diversity in different country settings will be confusing: "playing [this Cyberball] with other people from Serbia, for example... the majority of people, 99\%, are white people... thus, other groups like African Americans or Asian or whatever would be really unrealistic" (R9). 

\subsubsection*{Ball-Tossing Game and Background Environment Design}
Many naive participants expected the overall game and background design to be more realistic and detailed: "[the current design is] not that engaging and exciting since we could just see what the avatars are doing, and it is hard to understand the point of the process" (P2). For future iterations, P3 proposed that the throwing force can determine whether the ball reaches the other avatar, and the avatar could also miss the ball while catching. Plus, P1, P3, and P5 shared that they want a more detailed background, like adding visuals of grass, trees, and forest sounds (P3). P1 and P5 also envisioned having a "cloudy sky," "the sounds of wind-breezing and birds," and "people talking and cheers," etc. From their perspectives, a realistic background environment with audio help people feel more present. However, most researchers (R3, R4, R5, R6, R8, R9, and R10) strongly disagreed with this, mentioning the importance of keeping the design minimal for research. R3, R5, R9, and R10 expressed that any minor design differences make it more difficult to validate the findings. R8, R9, and R10 also reported concerns that additional visuals in the background may "distract people from the game."

\section{Discussion: Multi-Stakeholder Perspectives for VR Design}

\begin{figure}[h]
 \begin{center}
 \vspace{-9pt}
\includegraphics[width=0.48\textwidth]{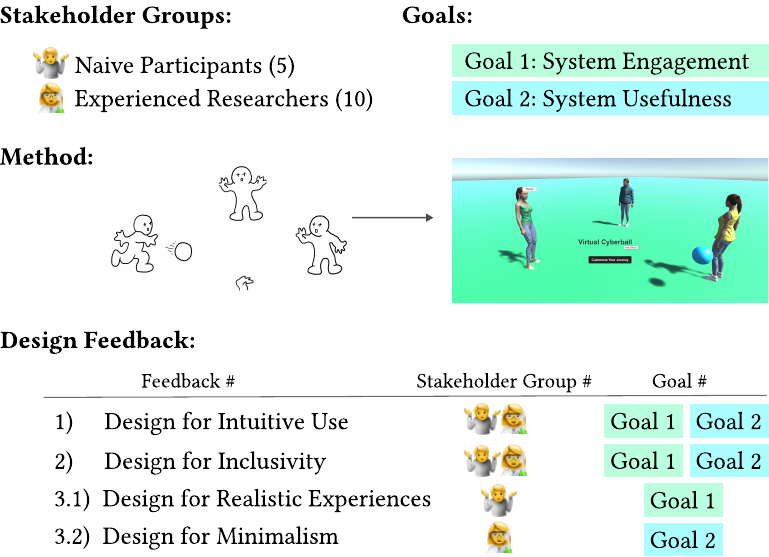}
\Description[An illustration of the alignments and misalignments of feedback and goals between the two stakeholder groups]{Shared design insights between the two groups are 1) design for intuitive use and 2) design for inclusivity, and these two insights are all supported by the two goals. However, for the 3) design for realistic experiences and design for minimalism, the two groups hold different opinions due to their different goal.}
\caption{The (mis-)alignments of feedback and goals between the two stakeholder groups}
 \end{center}
  \vspace{-13pt}
\end{figure}

This paper describes the design of a virtual version of the Cyberball paradigm and the testing process with two groups of stakeholders: naive participants and researchers. At first, the two groups agreed on the importance of designing a virtual Cyberball and shared satisfaction with it. They shared expectations for the system design to be more intuitive (4.1) and inclusive (4.2), recognizing that these enhancements can greatly enhance their interaction with the tool by making it more \textbf{\textit{engaging}} and fulfill their research needs by making the tool more \textbf{\textit{useful}} in the context of scientific study. Therefore, the agreed-upon feedback from the stakeholder groups will be incorporated into the next iteration of the system.

However, regarding the detailed game experience (4.3), they held different opinions. In general, naive participants, seeing this as a one-time experience, focused on the \textbf{\textit{engaging}} side: whether their interaction with the system was intriguing and fun. They mainly relied on their previous VR game experiences as a baseline without considering the bigger picture of "control" in the research setting. In comparison, researchers focus more on the \textbf{\textit{useful}} side and scientific objectivity: they envision a simple system that rules out the confounding elements while providing an immersive, embodied experience to test social exclusion. Though we cannot directly incorporate these diverging perspectives, these insights still offer designers two ways to consider how to design an interactive system for distinct stakeholder groups regarding interactivity, aesthetics, inclusivity, and experience design.

Previous multi-stakeholder perspectives (MSP) works in designing civic technology \cite{multistakeholderCivic}, urban planning \cite{multistakeholderArch}, medical intervention \cite{multistakeholderHealth}, and digital tools for sensitive user groups \cite{multistakeholderAsylum} all demonstrate the usefulness of  contextualizing design guidelines from various stakeholder groups due to their distinct backgrounds and needs. 

However, there are only a few MSP design studies for VR, and most are for teaching and medical usages \cite{msVRhighedu, msVRkidpdarkphobia, msHosmedical}. Thus, applying this method to designing future virtual Cyberballs and other virtual reality paradigms used for social science research is essential. Such implementations will need to take into account the engagement and attention of the naive participants, who will judge an environment based on their previous experience with games and other virtual environments. However, they will also need to accommodate the needs of social science researchers for experimental control and precision. In conclusion, we strongly recommend that future researchers and designers should integrate the MSP approach into their VR design and study processes.

Despite the potential, there still exist gaps in our study, which also open many future opportunities: 

1) Broader Sampling:  Our interviews highlighted the importance of providing equitable representation across several avatar characteristics, including age, gender, body type, and clothing options. This aligns with considerable other work finding disparities in avatar representation \cite{genderage, genderbelt, pace2009socially, chiEA}. Our small sample size indicated these categories as most salient, while future work should use wider sampling to determine challenges in avatar representation.

2) Embodiment Effects:  Embodiment is becoming a more commonly experienced aspect of interactions in virtual worlds~\cite{10.1145/3544548.3580918, latoschik2019not, smith2018communication}. Understanding how representations of the user's own body in the social science experimental platform becomes increasingly necessary: how do people perceive and interact with representations of their own bodies inside VR Cyberball and other social science paradigms? We hope our findings contribute toward this area. 

3) Limitations of the MSP Approach: Though dual- and multi-stakeholder perspectives are valuable, identifying and connecting with the stakeholder groups could be challenging \cite{cha}. Also, since every stakeholder has different personal background and technology experiences, intergroup disagreements are expected to occur frequently. Thus, we should address these concerns in future VR design using the MSP approach.

\section{Acknowledgements}
We express our sincere gratitude to all the researchers and participants who dedicated their time to participate in this study and provide valuable feedback. This work, being the first author's undergraduate research honors project, received generous support from the Jane E. Brody Undergraduate Research Award and the Fredric N. Gabler ’93 Memorial Research Endowment. Furthermore, the preliminary project of this work was recognized with research distinction at Cornell University in May 2022. 

\bibliographystyle{ACM-Reference-Format}
\bibliography{bib}

\end{document}